\documentstyle[aps,twoside]{revtex}
\begin{document}
\twocolumn[\hsize\textwidth\columnwidth\hsize\csname@twocolumnfalse%
\endcsname
\draft
\title{On QCD predictions for the chiral Lagrangian coefficients}
\vspace{0.2cm}
\author{Qing Wang$^{b,a}$,~~~~ 
Yu-Ping Kuang$^{a,b}$,~~~~Hua Yang$^{b,c}$,~~~~Qin Lu$^b$}
\vspace{0.2cm}
\address{
a. CCAST (World Laboratory),
P.O.Box 8730, Beijing 100080, China \\
b. Department of Physics, Tsinghua University, Beijing 100084, China
\footnote{Mailing address}\\
c. Institute of Electronic Technology, Information Engineering
University, Zhengzhou 450004, Henan, China}
\date{TUHEP-TH-02134}
\bigskip

\maketitle

\begin{abstract}
Based on a previous study of deriving the chiral Lagrangian (CL) from
QCD, we illustrate the main feature of QCD predictions for the CL 
coefficients (CLC) in certain approximations. We first 
show that, in the large-$N_c$ limit, the anomaly 
part contributions to the CLC are exactly cancelled by certain terms in the 
normal part contributions (NPC), 
so that the final results only concern the remaining NPC 
depending on QCD interactions.
We then do the calculation in a simple approach
with further approximations. The obtained CLC and quark condensate 
are consistent with the experiments.
\end{abstract}


\vspace{0.2cm}
]

\vspace{0.2cm} 
\section{INTRODUCTION}

Studying low energy hadron physics in QCD is a long standing difficult problem 
due to its nonperturbative nature. A widely used approach is the theory
of chiral Lagrangian (CL) based on the global symmetry of the system and the 
momentum expansion without dealing with the nonperturbative QCD dynamics 
\cite{weinberg}\cite{GL}. In such an approach, the CL coefficients (CLC) are 
all unknown parameters being determined by experimental inputs. 
Studying the relation 
between the CL and the underlying theory of QCD will not only be theoretically 
interesting for a deeper understanding of the CL, but will also
be helpful for increasing the predictive power of the CL.

The CLC are contributed both by the anomaly part (from the quark functional 
measure) and the normal part (from the QCD Lagrangian). 
In the literature, the anomaly part contributions (APC) are carefully 
calculated by using the heat kernel technique, 
while the normal part contributions (NPC) have not been calculated as 
carefully \cite{Espriu,Simic}.
In a previous paper, Ref.\cite{WKWX1}, the CL was formally derived from the 
first principles of QCD without taking approximations, and all the NPC to
the CLC are expressed in terms of certain Green's functions in QCD.
To compare the APC and NPC carefully, a unified regularization scheme and new 
technique feasible for the calculations of both APC and NPC is needed since 
the heat kernel technique is hard to be applied to the NPC which contains 
complicated functions of the momentum from nonperturbative QCD dynamics.

In this letter, we take the Schwinger proper time regularization scheme
for the calculations of both APC and NPC with the technique developed in 
\cite{det}. Thus APC and NPC are treated on equal footing.
We then take a simple approach with a series of approximations to 
illustrate the main feature of how QCD predicts the CLC. The first 
approximation is to take the {\it large-$N_c$ limit} in which the effective 
actions can be evaluated in the saddle-point approximation, and {\it all the 
CLC should be free from ultraviolet divergence} since the only
ultrviolet divergence in the CLC comes from the 
meson-loop corrections \cite{GL} which are of $O(1/N_c)$.

We first calculate the APC. The anomaly term in
the path integral can be expressed by the following effective action
\begin{eqnarray}                                      
S_{\rm eff}^{({\rm anom})}&\equiv&-i\times{\rm anomaly~~ terms}\nonumber\\
&=&
-iN_c[{\rm Tr}\ln(i\partial\!\!\!\! /\;+J)
-{\rm Tr}\ln(i\partial\!\!\!\! /\;+J_{\Omega})]\nonumber\\
&=&iN_c[{\rm Tr}\ln(i\partial\!\!\!\! /\;+J_{\Omega})
+\cdots],
\label{Sanom}
\end{eqnarray}
where $\Omega$ is related to the nonlinearly realized meson field $U$ by
$U=\Omega^2$, $J$ is the external source containing scalar,
pseudoscalar, vector, and axial vector components, $J_\Omega$ is 
$J$ chirally rotated by $\Omega$ \cite{WKWX1}, and the ellipsis denotes 
$\Omega$-independent ($U$-independent) terms which is irrelevant to the CLC. 
The APC to the CLC concerns only the real part of the
${\rm Tr}\ln(i\partial\!\!\!\! /\;+J_{\Omega})$ term in (\ref{Sanom}),
which is positively definite in the Euclidean space-time \cite{Ball}.
Now we evaluate it by using the Schwinger proper time regularization
with parameters $\Lambda$ and $\kappa$ regularizing the UV and IR 
divergences, respectively.
In \cite{WKWX1}, we see that the APC to the $O(p^2)$ CLC is exactly 
cancelled by a term in the NPC, so that we concentrate on the $O(p^4)$ CLC.
After lengthy but elementary calculations, we obtain
\begin{eqnarray}                          
&&L_1^{\rm (anom)}
=\frac{N_c}{384\pi^2},~
L_2^{\rm (anom)}
=\frac{N_c}{192\pi^2},~
L_3^{\rm (anom)}
=-\frac{N_c}{96\pi^2},\nonumber\\
&&L_4^{\rm (anom)}=L_5^{\rm (anom)}=L_6^{\rm (anom)}=0,~
L_7^{\rm (anom)}
=\frac{N_c}{1152\pi^2},\nonumber\\
&&L_8^{\rm (anom)}
=-\frac{N_c}{384\pi^2},~
L_9^{\rm (anom)}
=\frac{N_c}{48\pi^2},~
L_{10}^{\rm (anom)}
=-\frac{N_c}{96\pi^2}\nonumber,\\
&&H_1^{\rm (anom)}
=\frac{N_c}{96\pi^2}\lim_{\kappa\rightarrow 0}
\lim_{\Lambda\rightarrow\infty}(\ln\frac{\kappa^2}{\Lambda^2}
+\gamma+\frac{1}{2}),\nonumber\\
&&H_2^{\rm (anom)}
=\frac{N_c}{192\pi^2}+
\lim_{\Lambda\rightarrow\infty}\frac{N_c\Lambda^2}{32\pi^2B_0^2}.
\label{anomalyresults}
\end{eqnarray}
These are exactly the results in \cite{Espriu} for $M_Q=0$\footnote{
Different from the approach in
\cite{Espriu}, our present approach does not put in a constituent
quark mass $M_Q$ by hand. Instead, it is naturally included in the
quark self-energy reflecting chiral symmetry breaking in the normal
part. So that our results should be compared with the $M_Q=0$ results
in \cite{Espriu}.}. When taking $N_c=3$, the results given in 
(\ref{anomalyresults}) are close to the experimental results \cite{GL}
except that $L_7$ and $L_8$ are of wrong signs. This gives people an
impression that the APC might play the major role in the CLC, and the
NPC might only contribute small corrections \cite{Espriu,Simic}.
We argue that {\it the results in (\ref{anomalyresults}) should not
appear in the final expressions for the CLC} because: (i) they are 
independent of QCD interactions, while the final CLC describing the
interactions between mesons (residual interactions between quarks and gluons) 
should depend on QCD interactions, (ii) the final CLC should be finite
as $\Lambda\to\infty$ in the 
large-$N_c$ limit, while $H_1$ and $H_2$ in (\ref{anomalyresults}) are 
divergent as $\Lambda\to\infty$. We shall see later that {\it the terms in 
(\ref{anomalyresults}) are indeed exactly cancelled by certain terms in the
NPC, and they do not really appear in the final expressions for the CLC.}

Now, we calculate the NPC with the same regularization scheme. In the 
large-$N_c$ limit, the saddle-point approximation reduces the normal part 
effective action $S^{(\rm norm)}_{\rm eff}$ to
\begin{eqnarray}                                
&&S_{\rm eff}^{({\rm norm})}=
-iN_c{\rm Tr}\ln[i\partial\!\!\!/+J_{\Omega}-\Pi_{\Omega c}]
+N_c\int d^{4}xd^{4}x'\nonumber\\
&&\hspace{0.4cm}\times
\Phi^{\sigma\rho}_{\Omega c}(x,x')\Pi^{\sigma\rho}_{\Omega c}(x,x')
+N_c\sum^{\infty}_{n=2}{\int}d^{4}x_1\cdots d^4x_{n}'\nonumber\\
&&\hspace{0.4cm}\times
\frac{(-i)^{n}(N_c g_s^2)^{n-1}}{n!}\bar{G}^{\sigma_1\cdots\sigma_n}_{\rho_1
\cdots\rho_n}(x_1,x'_1,\cdots,x_n,x'_n)\nonumber\\
&&\hspace{0.4cm}\times\Phi^{\sigma_1\rho_1}_{\Omega c}(x_1 ,x'_1)\cdots 
\Phi^{\sigma_n\rho_n}_{\Omega c}(x_n ,x'_n)
\label{Snorm}
\end{eqnarray}
satisfying the useful relation \cite{WKWX1}
\begin{eqnarray}                           
\frac{d S_{\rm eff}^{({\rm norm})}}
{d J^{\sigma\rho}_{\Omega}(x)}
\bigg|_{U \mbox{ fix, anomaly ignored}}
=N_c\overline{\Phi_{\Omega c}
^{\sigma\rho}(x,x)}\label{Seffdiff},
\end{eqnarray}
where $\Phi$, and $\Pi$ are auxiliary fields, $\Phi_{\Omega c}$ and 
$\Pi_{\Omega c}$ are classical fields of the chirally rotated (by
$\Omega$) $\Phi$ and $\Pi$ satisfying the saddle-point equations, 
and $\bar{G}$ is the generalized gluon Green's function defined in 
\cite{WKWX1}. In the present approximation, (\ref{Seffdiff}) reduces to 
\begin{eqnarray}                            
-i[(i\partial\!\!\! /\;+J_{\Omega}
-\Pi_{\Omega c})^{-1}]^{\rho\sigma}(x,x)
=\Phi^{\sigma\rho}_{\Omega c}(x,x).
\end{eqnarray}
We see that $\Pi_{\Omega c}$ and $\Phi_{\Omega c}$ play the roles of
the quark self-energy and the quark propagator, respectively, in the
case with $J_\Omega\ne 0$.

Next, we decompose $S_{\rm eff}^{({\rm norm})}$ into a part
$S^{(\rm norm,\Pi_{\Omega c}=0)}_{\rm eff}$ {\it independent of
$\Pi_{\Omega c}$} and a part $S^{(\rm norm,\Pi_{\Omega c}\ne 0)}_{\rm
eff}$ {\it depending on $\Pi_{\Omega c}$}. 
$S^{(\rm norm,\Pi_{\Omega c}=0)}_{\rm eff}$
can be extracted from (\ref{Snorm}) by setting $\Pi_{\Omega c}=0$, i.e.
\begin{eqnarray}                            
&&S^{({\rm norm},\Pi_{\Omega c}=0)}_{\rm eff}
=-iN_c{\rm Tr}\ln[i\partial\!\!\!/+J_{\Omega}]\nonumber\\
&&\hspace{0.4cm}+N_c\bigg[\sum^{\infty}_{n=2}{\int}d^{4}x_1\cdots d^4x_{n}'
\frac{(-i)^{n}(N_c g_s^2)^{n-1}}{n!}\nonumber\\
&&\hspace{0.4cm}\times
\bar{G}^{\sigma_1\cdots\sigma_n}_{\rho_1
\cdots\rho_n}(x_1,x'_1,\cdots,x_n,x'_n)\nonumber\\
&&\hspace{0.4cm}\times\Phi^{\sigma_1\rho_1}_{\Omega c}(x_1 ,x'_1)\cdots 
\Phi^{\sigma_n\rho_n}_{\Omega c}(x_n ,x'_n)\bigg]_{\Pi_{\Omega c}=0}.
\label{SPi=0}
\end{eqnarray}
It can be shown that the last term in (\ref{SPi=0}) is actually 
$\Omega$-independent \cite{YWKL}. Therefore, (\ref{SPi=0}) can be written as 
\begin{eqnarray}                            
S^{({\rm norm},\Pi_{\Omega c}=0)}_{\rm eff}
=&&-iN_c[{\rm Tr}\ln(i\partial\!\!\!/+J_{\Omega})+\cdots].
\label{SPi=0'}
\end{eqnarray}
where the ellipsis denotes $\Omega$-independent terms irrelevant to the
CLC.
Comparing the $\Omega$-dependent terms in (\ref{Sanom}) and
(\ref{SPi=0'}), we see that they are of the same form but with an
opposite sign. Thus {\it their contributions to the CLC exactly cancel each 
other to all orders in the momentum expansion}.
Hence {\it the results in (\ref{anomalyresults}) do not really appear in
the final expressions for the CLC. The CLC are actually contributed by
the remaining normal part effective action 
$S_{\rm eff}^{({\rm norm,\Pi_{\Omega c}\ne 0})}$}.
{\it This is our first new conclusion in this study}. 
Then the final formulae for the CLC given in \cite{WKWX1} (cf.
eqs.(62) and (63) in \cite{WKWX1}) becomes
\begin{eqnarray}                          
&&L_1
=\frac{1}{32}\tilde{\cal K}_4
+\frac{1}{16}\tilde{\cal K}_5
+\frac{1}{16}\tilde{\cal K}_{13}
-\frac{1}{32}\tilde{\cal K}_{14}\nonumber,\\
&&L_2
=\frac{1}{16}(\tilde{\cal K}_4
+\tilde{\cal K}_6)+\frac{1}{8}\tilde{\cal K}_{13}
-\frac{1}{16}\tilde{\cal K}_{14}\nonumber,\\
&&L_3
=\frac{1}{16}(\tilde{\cal K}_3
-2\tilde{\cal K}_4
-6\tilde{\cal K}_{13}
+3\tilde{\cal K}_{14})\nonumber,\\
&&L_4
=\frac{\tilde{\cal K}_{12}}{16B_0}\nonumber,~~
L_5
=\frac{\tilde{\cal K}_{11}}{16B_0}\nonumber,~~
L_6
=\frac{\tilde{\cal K}_8}{16B_0^2}\nonumber,\\
&&L_7
=-\frac{\tilde{\cal K}_1}{16N_f}
-\frac{\tilde{\cal K}_{10}}{16B_0^2}
-\frac{\tilde{\cal K}_{15}}{16B_0N_f}\nonumber,\\
&&L_8
=\frac{1}{16}[\tilde{\cal K}_1+\frac{1}{B_0^2}\tilde{\cal K}_7
-\frac{1}{B_0^2}\tilde{\cal K}_9
+\frac{1}{B_0}\tilde{\cal K}_{15}]\nonumber,\\
&&L_9
= \frac{1}{8}(4\tilde{\cal K}_{13}
-\tilde{\cal K}_{14}),~~
L_{10}
=\frac{1}{2}(\tilde{\cal K}_2
-\tilde{\cal K}_{13})
\nonumber,\\
&&H_1
=-\frac{1}{4}(\tilde{\cal K}_2
+\tilde{\cal K}_{13})\nonumber,\\
&&H_2
=\frac{1}{8}[-\tilde{\cal K}_1
+\frac{1}{B_0^2}\tilde{\cal K}_7
+\frac{1}{B_0^2}\tilde{\cal K}_9
-\frac{1}{B_0}\tilde{\cal K}_{15}],
\label{p4full}
\end{eqnarray}
in which $\tilde{\cal K}_i\equiv {\cal K}_i^{(\rm
norm,\Pi_{\Omega c}\ne 0)}={\cal K}_i^{(\rm norm)}-{\cal K}_i^{(\rm
norm,\Pi_{\Omega c}=0)}$, and the ${\cal K}$s are components with
various Lorentz structures of the related QCD Green's functions defined
in \cite{WKWX1}. These $O(p^4)$ CLC depend on QCD interactions through
$\Pi_{\Omega c}$ as it should be.

The effective action $S^{(\rm norm)}_{\rm eff}$ in
(\ref{Snorm}) has never been carefully evaluated in the literature. As
the first time of doing the calculation, we take further
approximations to simplify the evaluation. Now we
take the approximation of {\it keeping only the leading order in
dynamical perturbation} \cite{PS}, which means taking account of only the
nonperturbative QCD dynamics through the quark self-energy reflecting
chiral symmetry breaking, and neglecting all QCD corrections in
positive powers of $g_s$. Thus the complicated last term in
(\ref{Snorm}) is neglected. Moreover, from the equation of motion of
$\Phi_{\Omega c}$, we can see that the second term is of the same order as
the last term so that it should be neglected as well. Then only the
first term in (\ref{Snorm}) is kept. As we have mentioned,
$\Pi_{\Omega c}$ plays the role of the quark self-energy. From the
local gauge transformation property of $\Pi_{\Omega c}$, we know that
$\Pi_{\Omega c}$ is related to the conventional quark self-energy
$\Sigma(-p^2)$ by
\begin{eqnarray}                      
\Pi^{\sigma\rho}_{\Omega c}(x,y)=
[\Sigma(\overline{\nabla}^2_x)]^{\sigma\rho}\delta^4(x-y),
\end{eqnarray}
where $\bar\nabla^\mu_x=\partial^\mu_x-iv^\mu_\Omega(x)$.
Then the simplified $S^{(\rm norm)}_{\rm eff}$ can be written as
\begin{eqnarray}                      
S_{\rm eff}^{({\rm norm})}
=-iN_c{\rm Tr}\ln[i\partial\!\!\!/+J_{\Omega}
-\Sigma(\overline{\nabla}^2)].
\label{SGL}
\end{eqnarray}
This can be evaluated in the Schwinger proper time regularization scheme
with the technique developed in \cite{det}. The calculation is
first performed in the Euclidean space-time and then converted into the
Minkowskian space-time. The calculation is lengthy and the final
results in the Minkowskian space-time are
\begin{eqnarray}             
&&F_0^2B_0=4\int d\tilde{p} \Sigma_pX_p,
\label{F0B0}\nonumber\\
&&F_0^2=2\!\int\!\!d\tilde{p}\bigg[(-2\Sigma^2_p\!-\!p^2\Sigma_p\Sigma'_p)X_p^2
\!+\!(2\Sigma^2_p
+p^2\Sigma_p\Sigma'_p)
\frac{X_p}{\Lambda^2}\bigg],
\label{F0}\nonumber\\
&&{\cal K}^{({\rm norm})}_{1}=2\int d\tilde{p}\bigg[-2A_pX_p^3
+2A_p\frac{X_p^2}{\Lambda^2}-A_p\frac{X_p}{\Lambda^4}
\nonumber\\
&&\hspace{0.4cm}+\frac{p^2}{2}\Sigma'^2_p\frac{X_p}{\Lambda^2}
-\frac{p^2}{2}\Sigma'^2_pX_p^2,
\bigg],
\nonumber\\
&&{\cal K}^{({\rm norm})}_{2}=\int d\tilde{p}\bigg[-2B_pX_p^3
 +2B_p\frac{X_p^2}{\Lambda^2}-B_p\frac{X_p}{\Lambda^4}\nonumber\\
&&\hspace{0.4cm}+\frac{p^2}{2}\Sigma^{\prime 2}_p\frac{X_p}{\Lambda^2}
-\frac{p^2}{2}\Sigma^{\prime 2}_pX_p^2\bigg],
\nonumber\\
&&{\cal K}^{({\rm norm})}_{3}=2\int d\tilde{p}\bigg[
(\frac{4\Sigma^4_p}{3}-\frac{2p^2\Sigma^2_p}{3}+\frac{p^4}{18})(
 6X_p^4-\frac{6X_p^3}{\Lambda^2}\nonumber\\
&&\hspace{0.4cm}+\frac{3X_p^2}{\Lambda^4}
-\frac{X_p}{\Lambda^6})+(-4\Sigma^2_p+\frac{p^2}{2})(-2X_p^3
+\frac{2X_p^2}{\Lambda^2}-\frac{X_p}{\Lambda^4})
\nonumber\\
&&\hspace{0.4cm}-\frac{X_p}{\Lambda^2}+X_p^2\bigg],
\nonumber\\
&&{\cal K}^{({\rm norm})}_{4}=\int d\tilde{p}\bigg[
(\frac{-4\Sigma^4_p}{3}+\frac{2p^2\Sigma^2_p}{3}+\frac{p^4}{18})(
6X_p^4-\frac{6X_p^3}{\Lambda^2}\nonumber\\
&&\hspace{0.4cm}+\frac{3X_p^2}{\Lambda^4}
-\frac{X_p}{\Lambda^6})
+4\Sigma^2_p(-2X_p^3+\frac{2X_p^2}{\Lambda^2}
-\frac{X_p}{\Lambda^4})+\frac{X_p}{\Lambda^2}
-X_p^2\bigg],
\nonumber\\
&&{\cal K}^{({\rm norm})}_5={\cal K}^{({\rm norm})}_6={\cal K}^{(\rm
norm)}_8={\cal K}^{(\rm norm)}_{10}={\cal K}^{(\rm norm)}_{12}=0,
\nonumber\\
&&{\cal K}^{({\rm norm})}_7=2\int d\tilde{p}\bigg[(3\Sigma^2_p+2p^2
\Sigma_p\Sigma'_p)X_p^2
+[-2\Sigma^2_p\nonumber\\
&&\hspace{0.4cm}-p^2(1+2\Sigma_p\Sigma'_p)]\frac{X_p}{\Lambda^2}\bigg],
\nonumber\\
&&{\cal K}^{({\rm norm})}_9=2\!\int \!\!d\tilde{p}\bigg[(\Sigma^2_p\!+\!
2p^2\Sigma_p
\Sigma'_p)X_p^2
\!-\!p^2(1\!+\!2\Sigma_p\Sigma'_p)
\frac{X_p}{\Lambda^2}\bigg],\nonumber
\end{eqnarray}
\begin{eqnarray}                  
&&{\cal K}^{({\rm norm})}_{11}=4\int d\tilde{p}\bigg[(-4\Sigma^3_p
+p^2\Sigma_p)X_p^3+(4\Sigma^3_p-p^2\Sigma_p)
\nonumber\\
&&\hspace{0.4cm}\times\frac{X^2_p}{\Lambda^2}-(2\Sigma^3_p
-\frac{1}{2}p^2\Sigma_p)\frac{X_p}{\Lambda^4}
+3\Sigma_p\frac{X_p}{\Lambda^2}
-3\Sigma_p X_p^2\bigg],
\nonumber\\
&&{\cal K}^{({\rm norm})}_{13}=\int d\tilde{p}\bigg[
(\frac{1}{3}p^2\Sigma'_p\Sigma''_p+\frac{1}{3}\Sigma_p\Sigma''_p)X_p
+(C_p\nonumber\\
&&\hspace{0.4cm}-D_p)\frac{X_p}{\Lambda^2}
-(C_p-D_p)X_p^2-2E_pX_p^3
+2E_p\frac{X_p^2}{\Lambda^2}\nonumber\\
&&\hspace{0.4cm}-E_p\frac{X_p}{\Lambda^4}\bigg],
\nonumber\\
&&{\cal K}^{({\rm norm})}_{14}=-4\int d\tilde{p}\bigg[
-2F_pX_p^3+2F_p\frac{X_p^2}{\Lambda^2}
-F_p\frac{X_p}{\Lambda^4}
\nonumber\\
&&\hspace{0.4cm}+\frac{p^2}{2}\Sigma_p^{\prime 2}\frac{X_p}{\Lambda^2}
-\frac{p^2}{2}\Sigma^{\prime 2}_pX_p^2\bigg],
\nonumber\\
&&{\cal K}^{({\rm norm})}_{15}=-4\int d\tilde{p}\bigg[
-(\Sigma_p+\frac{1}{2}p^2\Sigma'_p)\frac{X_p}{\Lambda^2}
+(\Sigma_p+\nonumber\\
&&\hspace{0.4cm}\frac{1}{2}p^2\Sigma'_p)
X_p^2\bigg],
\label{Kresult}
\end{eqnarray}
in which the short notations are
$\Sigma_p\equiv\Sigma(-p^2)$, $\int d\tilde{p}\equiv iN_c$$\int
\frac{d^4p}{(2\pi)^4}\exp{\{(p^2-\Sigma^2_p)/\Lambda^2\}}$, $X_p\equiv 
1/(p^2-\Sigma^2_p)$,
and $A_p.~B_p,~C_p,~D_p,~E_p,$ and $F_p$ are functions of $\Sigma_p$
with lengthy expressions given in \cite{YWKL}.

Taking $\Lambda\to\infty$ the expression for $F^2_0$ in (\ref{Kresult})
is just the well-known Pagels-Stokar formula \cite{PS}. It is easy to check 
that the ${\cal K}^{({\rm norm})}_i$ ($i=1,\cdots,15$) in
(\ref{Kresult}) do contain the $\Sigma_p$-independent 
($\Pi_{\Omega c}$-independent) piece mentioned
before which exactly cancel the APC in (\ref{anomalyresults}).
It can also be checked that the results in (\ref{Kresult}) are all
finite when $\Lambda\to\infty$ as it should be.

Subtracting the $\Sigma_p$-independent piece from (\ref{Kresult}), we
get the desired $\tilde{\cal K}_i(\Sigma_p)$ ($i=1,\cdots,15$) needed in
obtaining the $O(p^4)$ CLC in (\ref{p4full}).

The final step of the calculation is to calculate $\Sigma(p^2)$ from
the Schwinger-Dyson equation. We further take the {\it improved ladder 
approximation} as in the literature,
\begin{eqnarray}                      
\Sigma(p^2)-\frac{3N_c}{2}\int\frac{d^4q}{4\pi^3}
\frac{\alpha_s[(p-q)]}{(p-q)^2}
\frac{\Sigma(q^2)}{q^2+\Sigma^2(q^2)}
=0.
\label{eq0}
\end{eqnarray}
So far we only know the large momentum behavior of the running coupling
constant $\alpha_s(p-q)$. The low momentum behavior of it is not know
yet due to the ignorance of nonperturbative QCD dynamics. Inevitably, we have 
to take certain QCD motivated model for it as in the literature. We shall take 
the following Model A from \cite{Aoki}, and Model B and Model C from 
\cite{Munczek} as examples to do the calculation. 
They are
\vspace{0.8cm}
\begin{eqnarray}                      
{\rm A}:~\alpha_s(p)&&=7\frac{12\pi}{(33-2N_f)},
\hspace{0.2cm} \mbox{for}~\ln(p^2/\Lambda_{QCD}^2)\leq -2;\nonumber\\
&&=\{7-\frac{4}{5}[2+\ln(p^2/\Lambda_{QCD}^2)]^2\}
\frac{12\pi}{(33-2N_f)},
\nonumber\\
&&\hspace{0.8cm}\mbox{for}~-2\leq\ln(p^2/\Lambda_{QCD}^2)\leq 0.5;
\nonumber\\
&&=\displaystyle\frac{1}{\ln(p^2/\Lambda^2_{QCD})}\frac{12\pi}{(33-2N_f)}
,\nonumber\\
&&\hspace{0.8cm}\mbox{for}~0.5\leq\ln(p^2/\Lambda_{QCD}^2).
\label{omega1}\nonumber\\
{\rm B}:~\alpha_s(p)&&=  4\pi^3\eta^2p^2
\delta^4(p)+\frac{12\pi}{(33-2N_f)}
\frac{1}{\ln(2+p^2/\Lambda^2_{QCD})};
\label{omega2}\nonumber\\
{\rm C}:~\alpha_s(p)&&= 
\frac{4\pi^3}{\mu^2}p^2e^{-p^2/p_0^2}+
\frac{12\pi}{(33-2N_f)}\frac{1}{\ln(2+p^2/\Lambda^2_{QCD})}.
\label{omega3}\nonumber
\end{eqnarray}
We take the original values $\Lambda_{QCD}=484$ MeV (Model A),
$\Lambda_{QCD}=230$ MeV (Models B and C), and $p_0=380$ MeV (Model C), and 
determine other parameters by taking $F_0=93$ MeV as input\footnote{In the 
large-$N_c$ limit, there is no meson-loop correction to $F^2_0$. Therefore, 
$F_0$ should be identified with $f_\pi=93$ MeV.}. We further take the usual
approximation $\alpha_s(p-q)\approx \theta(p^2-q^2)\alpha_s(p^2)
+\theta(q^2-p^2)\alpha_s(q^2)$ with which the integral equation
(\ref{eq0}) can be converted into differential equation which is easy
to solve numerically. We have found the numerical solutions with the
desired asymptotic behavior
$\Sigma(p^2)\stackrel{p^2\to\infty}\longrightarrow
[\ln((p^2/\Lambda^2_{QCD})]^{\gamma-1}/p^2$
($\gamma\equiv (9N_c)/(2(33-2N_f))$)
characterizing chiral symmetry breaking for the three models. Then we
obtain the values of the $O(p^4)$ CLC from (\ref{p4full}) and
(\ref{Kresult}) which are
listed in TABLE I together with the experimental values \cite{GL} for 
comparison.
We see from TABLE I that: {\it (a) the obtained CLC are not so sensitive
to the forms of $\alpha_s(p)$, 
(b) all $L_1,\cdots,L_{10}$ are of the right orders of magnitude and the right 
signs, 
(c) the consistency with the experiments of $L_1,~L_2,~L_4,~L_6,$ and $L_{10}$ 
is at $1\sigma$ level, and that of  $L_3,~L_5,~L_7$ and $L_8$ is at 
$2\sigma$ level, 
(d) only $L_9$ deviates from
the experimental value by $(3\--4)\sigma$. }
Considering the large theoretical uncertainty in this simple approach, 
{\it the obtained $L_1,\cdots,L_{10}$ are all consistent with the 
experiments}. 

In addition to $L_1,\cdots,L_{10}$, we can also calculate the quark condensate
from the $O(p^2)$ CLC $F^2_0B_0$ with the relation in the present
approximation $\langle\overline{\psi}\psi\rangle=-N_fF_0^2B_0$ \cite{WKWX1}.
This is divergent when taking $\Lambda\to\infty$, so that it needs to be
renormalized. We take the renormalization counter term such that $\Lambda$ is 
replaced by a scale parameter $\mu$. The values of the renormalized quark 
condensate at $\mu=1$ GeV for the three models are
\begin{eqnarray}            
&&{\rm A}:~~~~~~~~~~\langle\bar\psi\psi\rangle_r=-(296~\mbox{MeV})^3,
\nonumber\\
&&{\rm B}:~~~~~~~~~~\langle\bar\psi\psi\rangle_r=-(296~\mbox{MeV})^3,
\nonumber\\
&&{\rm C}:~~~~~~~~~~\langle\bar\psi\psi\rangle_r=-(301~\mbox{MeV})^3.\nonumber
\label{condensaterenorm}
\end{eqnarray}
Considering the large theoretical uncertainty in this calculation, the obtained
$\langle\bar{\psi}\psi\rangle_r$ is also consistent with the
experimentally determined value 
$\langle\bar\psi\psi\rangle_{expt}=-(250~\mbox{MeV})^3$ from the QCD
sum rule at the scale of the typical haronic mass \cite{SVZ}.

In conclusion, we have calculated the CLC from QCD in certain approximations. 
We have first shown {\it the exact cancellation between the APC and the
$\Pi_{\Omega c}$-independent part of the NPC, so the final results of
the CLC concern only the $\Pi_{\Omega c}$-dependent part of the NPC}.
{\it Our obtained CLC and quark condensate are all consistent with the 
experiments}. Although the present approximations are rather crude, it does 
{\it reveal the main feature of how QCD predicts the CLC}. We see that
{\it the quark self-energy reflecting chiral symmetry breaking plays an
important role in the predictions for the CLC}. Study on improved
approximations is in progress.

This work is supported by the National Natural science Foundation of China, 
the Foundation of Fundamental Research grant of  Tsinghua University,
and a special grant from the Ministry of Education of China.  

\null

\onecolumn
\begin{table}[h]
\null\noindent
{\small{\bf TABLE I}. The obtained values of the $O(p^4)$ CLC (in units
of $10^{-3}$) for Model A, B, and C with $\Lambda\to\infty$ together with the
experimental values \cite{GL} for comparison.}
\begin{tabular}{c  c c c c c c c c c c}
 &$L_1$ &$L_2$ &$L_3$ &$L_4$ & $L_5$ & $L_6$ & $L_7$ &
$L_8$ & $L_9$ & $L_{10}$\\
\hline
A: & 1.10 &2.20 &-7.82 &0 &1.62 & 0 & -0.70&
1.75&5.07&-7.06\\ 
B: & 0.921 & 1.84 & -6.73 & 0 & 1.43 & 0 & -0.673 & 1.64 & 3.80 & 
-6.22\\ 
C: & 0.948 & 1.90 & -6.90 & 0 & 1.29 & 0 & -0.632 & 1.56 & 
3.95 & -6.21 \\
Expt &$0.9\pm 0.3$&$1.7\pm 0.7$&$-4.4\pm 2.5$&$0\pm 0.5$&$2.2\pm
0.5$&$0\pm 0.3$&$-0.4\pm 0.15$&$1.1\pm 0.3$&$7.4\pm 0.7$&$-6.0\pm 0.7$\\
\end{tabular}
\end{table}
\end{document}